\UseRawInputEncoding
\documentclass[%
 aip,
 jmp,%
 amsmath,amssymb,
]{revtex4-1}

\usepackage{graphicx}
\usepackage{dcolumn}
\usepackage{bm}

\begin{document}


\title[Basic postulates of some coordinate transformations]{Basic postulates of some coordinate transformations within material media}

\author{Zi-Hua Weng}
 \email{xmuwzh@xmu.edu.cn.}
\affiliation{
School of Aerospace Engineering, Xiamen University, Xiamen 361005, China\\
College of Physical Science and Technology, Xiamen University, Xiamen 361005, China
}%


\date{\today}

\begin{abstract}
  The paper aims to explore the physical quantities of several invariants, including the basic postulates of some types of crucial coordinate transformations, conservation laws and continuity equations, in the electromagnetic and gravitational fields. J. C. Maxwell first utilized the quaternions to describe the electromagnetic theory. Subsequent scholars make use of the octonions to study the physical properties of electromagnetic and gravitational fields simultaneously, including the octonion field strength, field source, angular momentum, torque and force. When an octonion coordinate system transforms rotationally, the scalar part of one octonion will remain unchanged, although the vector part of the octonion may alter. In the octonion space $\mathbb{O}$ , some invariants can be derived from this octonion property. A part of these invariants can be selected as the basic postulates of Galilean transformation or Lorentz transformation. Similarly, it is able to derive several invariants from the octonion property, in the transformed octonion space $\mathbb{O}_u$ . And that the invariants can be chosen as the basic postulates of a few new types of coordinate transformations. Further, the combination of invariants in the octonion spaces can be applied as the basic postulates of some new coordinate transformations, relevant to the norm of physical quantities. Through the analysis, it is easy to find that each conserved quantity has its preconditions from the perspective of octonion spaces. This is helpful to deepen the further understanding of the physical properties of conservation laws and other invariants.
\end{abstract}


\maketitle


\section{\label{sec:level1}Introduction}

Is there any new coordinate transformation, besides the Galilean transformation and Lorentz transformation? Why do you choose the Galilean transformation first in most cases? Can one conservation law (or continuity equation) be established unconditionally? For a long time, these simple and important issues have puzzled and attracted scholars. It was not until the emergence of the octonion field theory (short for the field theory described with the octonions) that these questions were answered partially. When the octonion coordinate system transforms rotationally, the scalar part of one octonion will remain unchanged. By means of this octonion property, it is able to achieve a few invariants. The combination of these invariants can be selected as the basic postulates for the Galilean transformation, Lorentz transformation and several new coordinate transformations, in the electromagnetic and gravitational fields. It expands our understanding of the simultaneity of invariants, conservation laws and continuity equations.

In 1756, M. V. Lomonosov first proposed the law of mass conservation. In 1777, A.-L. de Lavoisier verified the law of mass conservation again. Since then, the law of mass conservation \cite{treptow} has been accepted. On the other hand, the further study of mass led scholars to introduce the concepts of gravitational mass and inertial mass, respectively. These extensions should be taken into account by the law of mass conservation. The coverage of the mass concept should be further expanded.

In 1842, R. Mayer proposed the law of energy conservation. J. P. Joule studied the law of energy conservation, and measured the heat equivalent of work. In 1847, H. Helmholtz describes the law of energy conservation strictly. The number of energy terms has been increasing over time. These variations should be taken into account by the law of energy conservation. The scope of the energy concept should be further expanded \cite{josset}.

The scholars owe B. Franklin the creation of the law of charge conservation. In 1747, B. Franklin first mentioned the law of charge conservation. After that, the scholars assume that the law of charge conservation is true on both macro and micro scales \cite{jungmann,ignatiev}. In a vacuum, the electric charges are quantized. In one material medium \cite{FD1,FD2}, the material may possess a fractional electric charge \cite{FD3,FD4}. The influence of material media should be considered in the law of charge conservation \cite{gratus}.

The above analysis shows that the existing field theories have a few defects in the exploration of some invariants, conservation laws and continuity equations, relevant to the electromagnetic and gravitational fields.

(1) Coordinate transformations. The existing coordinate transformations mainly deal with the Galilean transformation and Lorentz transformation. The Galilean transformation involves the speed of light and radius vector. It believes that the time and mass both are invariable, and this is fit for the low-speed movement cases. The Lorentz transformation concerns the speed of light and norm of radius vector. It deems that the time and mass both are not invariable, and this is applied to explaining the high-speed movement cases. But these two coordinate transformations are unable to explore some invariants related to the electric charges.

(2) Unconstrained establishment. The existing field theory reckons that the electromagnetic field and gravitational field both are in the same space. It presupposes that all of conservation laws can be established simultaneously, while all continuity equations can be effective simultaneously. The conservation laws and continuity equations can also be valid simultaneously. And even the law of charge conservation and the law of mass conservation can be available simultaneously. However, the viewpoint leads to its inability to explain why the charge-to-mass ratio changes.

(3) Unvarying speed of light. According to the existing experiences and imaginations, the classical field theory takes it for granted that the speed of light is constant in a vacuum. Furthermore, this view has merely been verified in a quite limited number of experiments. Nevertheless the speed of light is variable in a large number of physical experiments. The existing studies have not fully considered the influence of various external factors on the speed of light. Apparently, this point of view limits the scope of application of the coordinate transformations.

In a stark contrast to the above, the octonion field theory is able to solve the puzzlement of some invariants, such as the simultaneous establishment of a few invariants, in the electromagnetic and gravitational fields. Also it can figure out some difficult problems derived from the existing field theories. By means of the octonion field theory, it is capable of exploring the physical properties of invariants, including the basic postulates of coordinate transformations, conservation laws and continuity equations.

J. C. Maxwell first applied not only the algebra of quaternions but also the vector analysis to explore the physical properties of electromagnetic fields. The subsequent scholars \cite{mironov} utilize the quaternions \cite{morita} and octonions \cite{deleo} to study the electromagnetic fields \cite{tanisli1,demir}, gravitational fields \cite{rawat}, invariants \cite{kansu1,kansu2}, conservation laws, continuity equations, quantum mechanics \cite{gogberashvili,bernevig}, relativity \cite{moffat}, astrophysical jets, weak nuclear fields \cite{majid,farrill}, strong nuclear fields \cite{furey,chanyal2}, black holes \cite{bossard}, and dark matters \cite{furui} and so forth.

In the paper, making use of the octonion field theory, it is able to explore some invariants, conservation laws and continuity equations, relevant to the electromagnetic and gravitational fields. And that the studies described with the octonion field theory possess a few advantages as follows.

(1) Norm of physical quantity. The transformed octonion space $\mathbb{O}_u$ is independent from the octonion space $\mathbb{O}$ (in Section 3). The Galilean transformation and Lorentz transformation in the octonion space $\mathbb{O}$ are the same as those in the existing field theory, respectively. The Lorentz transformation deals with the norm of octonion radius vector. Meanwhile, there are two new types of coordinate transformations, which are similar to the Galilean transformation or Lorentz transformation respectively, in the transformed octonion space $\mathbb{O}_u$ . Further, the norms of other physical quantities can also participate in the formation of a few new coordinate transformations, in particular the norm of octonion velocity.

(2) Restrictive establishment. The invariants in the transformed octonion space $\mathbb{O}_u$ cannot be established simultaneously with those in the octonion space $\mathbb{O}$ . As a result, the conservation laws are divided into two different types of groups, and the two groups of conservation laws are unable to be effective simultaneously. Next, the continuity equations will be separated into two different sets, while the two sets of continuity equations cannot be established simultaneously. In particular, the law of mass conservation and the law of charge conservation cannot be effective simultaneously. One of its direct inferences is that the charge-to-mass ratio must be varied. Moreover, there are many factors affecting the charge-to-mass ratio.

(3) Varying speed of light. The invariable speed of light is merely a choice. This assumption is only applicable to a few physical phenomena. Meanwhile the speed of light and even the optical refractive index are variable, in a large number of physical phenomena. That is, the speed of light is not an invariant in general. In the paper, only the Galilean transformation or Lorentz transformation can choose that the speed of light is an invariant. However, the other three major coordinate transformations, in the paper, have to select that the speed of light is not a constant. Apparently, various types of combinations of some invariants expand the scope of application of major coordinate transformations.

In the paper, when an octonion coordinate system transforms rotationally, the scalar part of one octonion will keep the same, although the vector part of the octonion may vary. As a result, a few invariants can be derived from this octonion property. By means of these invariants in the octonion space $\mathbb{O}$, it is able to achieve the basic postulates of Galilean transformation and Lorentz transformation. Similarly, it is capable of inferring the basic postulates of several new coordinate transformations in the transformed octonion space $\mathbb{O}_u$. Further, the combination of invariants may be considered as the basic postulates of some coordinate transformations, relevant to the norm of physical quantities. Through the above analysis and comparison, it is found that the Galilean transformation must be preferred in most cases, in order to obtain as many details of theoretical explanation as possible. This helps us to further understand the invariants, conservation laws and continuity equations.

\begin{table}[h]
\caption{Some equations relevant to the electromagnetic and gravitational fields in the octonion space. These octonion composite physical quantities consider the contributions of material media.}
\begin{ruledtabular}
\begin{tabular}{ll}
composite physical quantity~~~~~~     &  octonion definition                                                                                                    \\
\hline
composite field source                & $\mu \mathbb{S}^+ = - ( i \mathbb{F}^+ / v_0 + \lozenge )^\ast \circ \mathbb{F}^+$                                      \\
composite linear momentum             & $\mathbb{P}^+ = \mu \mathbb{S}^+ / \mu_g$                                                                               \\
composite property equation           & $\mathbb{P}^+ = k_{pf} ( \mathbb{F}^+ + \mathbb{F}_{ext}^+ )$                                                           \\
composite angular momentum            & $\mathbb{L}^+ = ( \mathbb{R} + k_{rx} \mathbb{X} )^\times \circ \mathbb{P}^+$                                           \\
composite torque                      & $\mathbb{W}^+ = - v_0 ( i \mathbb{F}^+ / v_0 + \lozenge ) \circ \{ ( i \mathbb{V}^\times / v_0 ) \circ \mathbb{L}^+ \}$ \\
composite force                       & $\mathbb{N}^+ = - ( i \mathbb{F}^+ / v_0 + \lozenge ) \circ \{ ( i \mathbb{V}^\times / v_0 ) \circ \mathbb{W}^+ \}$     \\
\end{tabular}
\end{ruledtabular}
\end{table}

\section{Octonion spaces}

R. Descartes believed that the space is the extension of substance. Nowadays, the Cartesian thought is improved to that the fundamental space is the extension of fundamental field \cite{weng1}. The fundamental fields include the gravitational field and electromagnetic field. Each fundamental field possesses one fundamental space. Each fundamental space is one quaternion space.

W. R. Hamilton invented the algebra of quaternions in 1843. Subsequently, J. T. Graves and A. Cayley discovered the octonion independently. The latter is called as the classical octonion. Besides, the scholars have proposed some other types of octonions. In this paper, the classical octonions are utilized to explore the physical properties of electromagnetic and gravitational fields.

J. C. Maxwell first applied the quaternion spaces to describe the physical properties of electromagnetic fields. Subsequent scholars used the quaternion spaces to study the electromagnetic theory or gravitational theory. Further, two independent quaternion spaces can be combined together to become one octonion space. The present scholars study the physical properties of electromagnetic field and gravitational field simultaneously by means of the octonion spaces.

In the octonion space $\mathbb{O}$ for the electromagnetic and gravitational fields, $\textbf{i}_j$ and $\textbf{I}_j$ both are the basis vectors, while $r_j$ and $R_j$ both are the coordinate values. The octonion radius vector is, $\mathbb{R} = \mathbb{R}_g + k_{eg} \mathbb{R}_e$ . The octonion velocity is,
$\mathbb{V} = \mathbb{V}_g + k_{eg} \mathbb{V}_e$. Herein $k_{eg}$ is one coefficient, to meet the demand for dimensional homogeneity. $\mathbb{R}_g = i \textbf{i}_0 r_0 + \Sigma \textbf{i}_k r_k$. $\mathbb{R}_e = i \textbf{I}_0 R_0 + \Sigma \textbf{I}_k R_k$. $\mathbb{V}_g = i \textbf{i}_0 v_0 + \Sigma \textbf{i}_k v_k$. $\mathbb{V}_e = i \textbf{I}_0 V_0 + \Sigma \textbf{I}_k V_k$. $r_j$ , $v_j$ , $R_j$ , and $V_j$ are all real. $r_0 = v_0 t$. $v_0$ is the speed of light, and $t$ is the time. $\textbf{i}_0 = 1$. $\textbf{i}_k^2 = -1$ . $\textbf{I}_j^2 = -1$ . $\textbf{I}_k = \textbf{i}_k \circ \textbf{I}_0$ . $\circ$ denotes the octonion multiplication. $i$ is the imaginary unit. $j = 0, 1, 2, 3. ~k = 1, 2, 3$.

In the octonion space $\mathbb{O}$ , the octonion field strength is $\mathbb{F}$ , the octonion field source is $\mathbb{S}$ , and the octonion linear momentum is $\mathbb{P}$ . From these octonion physical quantities, it is able to define the octonion angular momentum $\mathbb{L}$, torque $\mathbb{W}$ and force $\mathbb{N}$ . Further, the octonion field strength $\mathbb{F}$ and angular momentum $\mathbb{L}$ can be combined together to become the octonion composite field strength, $\mathbb{F}^+ = \mathbb{F} + k_{fl} \mathbb{L}$ . That is, $\mathbb{F}^+$ is the octonion field strength of electromagnetic and gravitational fields within the material media. Herein $k_{fl} = - \mu_g$ , which is the coefficient to satisfy the needs of dimensional homogeneity \cite{weng2}. $\mu_g$ is the gravitational constant.

In the octonion space $\mathbb{O}$ considering the contribution of material media, the octonion field source can be defined as, $\mu \mathbb{S}^+ = - ( i \mathbb{F}^+ / v_0 + \lozenge )^\ast \circ \mathbb{F}^+ $, within material media. The octonion linear momentum is defined as, $\mathbb{P}^+ = \mu \mathbb{S}^+ / \mu_g$ , within material media. The gravitational strength is, $\mathbb{F}_g^+ = \textbf{i}_0 f_0^+ + \Sigma \textbf{i}_k f_k^+$ , within material media. The electromagnetic strength is, $\mathbb{F}_e^+ = \textbf{I}_0 F_0^+ + \Sigma \textbf{I}_k F_k^+$ , within material media. The gauge condition is chosen as, $f_0^+ = 0$ and $F_0^+ = 0$, within material media. The octonion field source can be rewritten as, $\mu \mathbb{S}^+ = \mu_g \mathbb{S}_g^+ + \mu_e \mathbb{S}_e^+ - ( i \mathbb{F}^+ / v_0 )^\ast \circ \mathbb{F}^+$ , within material media. The gravitational source is, $\mathbb{S}_g^+ = i \textbf{i}_0 s_0^+ + \Sigma \textbf{i}_k s_k^+$ , within material media. The electromagnetic source is, $\mathbb{S}_e^+ = i \textbf{I}_0 S_0^+ + \Sigma \textbf{I}_k S_k^+$, within material media. Herein $f_0^+$ , $s_j^+$ , $F_0^+$ , and $S_j^+$ are all real. $f_k^+$ and $F_k^+$ both are complex numbers. $\mu$ is one coefficient. $\mu_e$ is the electromagnetic constant. $k_{eg}^2 = \mu_g / \mu_e$ . The quaternion operator is, $\lozenge = i \textbf{i}_0 \partial_0 + \Sigma \textbf{i}_k \partial_k$ , with $\partial_j = \partial / \partial r_j$ . $\ast$ denotes the octonion conjugate.

In the octonion composite property equation, there exists one relationship between the octonion composite linear momentum, $\mathbb{P}^+$ , with the octonion composite field strength, $( \mathbb{F}^+ + \mathbb{F}_{ext}^+ )$. The term, $\mathbb{F}_{ext}^+$ , is the octonion composite field strength from the external of material media. And $k_{pf}$ is one coefficient, to meet the demand for dimensional homogeneity (Table 1).

The octonion angular momentum within material media is defined as, $\mathbb{L}^+ = ( \mathbb{R} + k_{rx} \mathbb{X} )^\times \circ \mathbb{P}^+$ , with $\mathbb{X}$ being the octonion integrating function of field potential. The octonion composite angular momentum can be rewritten as, $\mathbb{L}^+ = \mathbb{L}_g^+ + k_{eg} \mathbb{L}_e^+$ , within material media. Herein $\mathbb{L}_g^+ = L_{10}^+ + i \textbf{L}_1^{i+} + \textbf{L}_1^+$ . $\mathbb{L}_e^+ = \textbf{L}_{20}^+ + i \textbf{L}_2^{i+} + \textbf{L}_2^+$ . The coefficient, $k_{rx} = 1 / v_0$ , is able to meet the demand for dimensional homogeneity. $\textbf{L}_1^+$ is the angular momentum within material media. $\textbf{L}_1^{i+}$ is called as the mass moment temporarily. $\textbf{L}_2^{i+}$ is the electric moment within material media, while $\textbf{L}_2^+$ is the magnetic moment within material media. $\textbf{L}_1^+ = \Sigma L_{1k}^+ \textbf{i}_k$ , $\textbf{L}_1^{i+} = \Sigma L_{1k}^{i+} \textbf{i}_k$. $\textbf{L}_2^+ = \Sigma L_{2k}^+ \textbf{I}_k$ , $\textbf{L}_2^{i+} = \Sigma L_{2k}^{i+} \textbf{I}_k$ . $\textbf{L}_{20}^+ = \textbf{I}_0 L_{20}^+$ . $L_{1j}^+$ , $L_{2j}^+$ , $L_{1k}^{i+}$ , and $L_{2k}^{i+}$ are all real. $\times$ represents the complex conjugate.

The octonion torque is, $\mathbb{W}^+ = - v_0 ( i \mathbb{F}^+ / v_0 + \lozenge ) \circ \{ ( i \mathbb{V}^\times / v_0 ) \circ \mathbb{L}^+ \}$, within material media. And it can be rewritten as, $\mathbb{W}^+ = \mathbb{W}_g^+ + k_{eg} \mathbb{W}_e^+$ . Herein $\mathbb{W}_g^+ = i W_{10}^{i+} + W_{10}^+ + i \textbf{W}_1^{i+} + \textbf{W}_1^+$. $\mathbb{W}_e^+ = i \textbf{W}_{20}^{i+} + \textbf{W}_{20}^+ + i \textbf{W}_2^{i+} + \textbf{W}_2^+$ . $W_{10}^{i+}$ is the energy within material media. $\textbf{W}_1^{i+}$ is the torque within material media, including the gyroscopic torque. $\textbf{W}_{20}^{i+}$ is called as the second-energy within material media temporarily. $\textbf{W}_2^{i+}$ is called as the second-torque within material media temporarily. $\textbf{W}_1^+ = \Sigma W_{1k}^+ \textbf{i}_k$ , $\textbf{W}_1^{i+} = \Sigma W_{1k}^{i+} \textbf{i}_k$ . $\textbf{W}_2^+ = \Sigma W_{2k}^+ \textbf{I}_k$ , $\textbf{W}_2^{i+} = \Sigma W_{2k}^{i+} \textbf{I}_k$ , $\textbf{W}_{20}^{i+} = W_{20}^{i+} \textbf{I}_0$ . $\textbf{W}_{20}^+ = W_{20}^+ \textbf{I}_0$ . $W_{1j}^+$ , $W_{2j}^+$ , $W_{1j}^{i+}$ , and $W_{2j}^{i+}$ are all real.

The octonion force is, $\mathbb{N}^+ = - ( i \mathbb{F}^+ / v_0 + \lozenge ) \circ \{ ( i \mathbb{V}^\times / v_0 ) \circ \mathbb{W}^+ \}$, within material media. And it will be rewritten as, $\mathbb{N}^+ = \mathbb{N}_g^+ + k_{eg} \mathbb{N}_e^+$ . Herein $\mathbb{N}_g^+ = i N_{10}^{i+} + N_{10}^+ + i \textbf{N}_1^{i+} + \textbf{N}_1^+$. $\mathbb{N}_e^+ = i \textbf{N}_{20}^{i+} + \textbf{N}_{20}^+ + i \textbf{N}_2^{i+} + \textbf{N}_2^+$ . $N_{10}^+$ is the power within material media. $\textbf{N}_1^{i+}$ is the force within material media, including the Magnus force. $\textbf{N}_{20}^{i+}$ is called as the second-power within material media temporarily. $\textbf{N}_2^{i+}$ is called as the second-force within material media temporarily. $\textbf{N}_1^+ = \Sigma N_{1k}^+ \textbf{i}_k$ , $\textbf{N}_1^{i+} = \Sigma N_{1k}^{i+} \textbf{i}_k$ . $\textbf{N}_2^+ = \Sigma N_{2k}^+ \textbf{I}_k$, $\textbf{N}_2^{i+} = \Sigma N_{2k}^{i+} \textbf{I}_k$ , $\textbf{N}_{20}^{i+} = N_{20}^{i+} \textbf{I}_0$. $\textbf{N}_{20}^+ = N_{20}^+ \textbf{I}_0$ . $N_{1j}^+$ , $N_{2j}^+$ , $N_{1j}^{i+}$ , and $N_{2j}^{i+}$ are all real.

Through the analysis and comparison, it can be found that there are some invariants in the octonion space $\mathbb{O}$. These invariants mainly relate to the physical quantities in the gravitational fields. On the other hand, there are some other invariants in the transformed octonion space $\mathbb{O}_u$ . These invariants mainly involve with the physical quantities in the electromagnetic fields. These two octonion spaces, $\mathbb{O}$ and $\mathbb{O}_u$ , are independent of each other. The invariants in the octonion spaces $\mathbb{O}$ are incompatible with those in the transformed octonion space $\mathbb{O}_u$ .

In the octonion space $\mathbb{O}$ , when any material medium does not make a contribution to the physical quantities, it is able to define some octonion physical quantities, including the octonion field strength $\mathbb{F}$, field source $\mathbb{S}$, linear momentum $\mathbb{P}$, angular momentum $\mathbb{L}$, torque $\mathbb{W}$, and force $\mathbb{N}$. Further, when the material media make a contribution to the physical quantities, we can define some more practical octonion physical quantities, including the octonion composite field strength $\mathbb{F}^+$, field source $\mathbb{S}^+$, linear momentum $\mathbb{P}^+$, angular momentum $\mathbb{L}^+$, torque $\mathbb{W}^+$, and force $\mathbb{N}^+$.

Similarly, in the transformed octonion space $\mathbb{O}_u$, when any material medium does not make a contribution to the physical quantities, it is capable of defining several transformed octonion physical quantities, including the octonion field strength $\mathbb{F}_u$, field source $\mathbb{S}_u$, linear momentum $\mathbb{P}_u$, angular momentum $\mathbb{L}_u$, torque $\mathbb{W}_u$, and force $\mathbb{N}_u$. Next, when the material media make a contribution to the physical quantities, we may define some more practical octonion physical quantities, including the octonion composite field strength $\mathbb{F}^+_u$, field source $\mathbb{S}^+_u$, linear momentum $\mathbb{P}^+_u$, angular momentum $\mathbb{L}^+_u$, torque $\mathbb{W}^+_u$, and force $\mathbb{N}^+_u$.

It is capable of deriving a few coordinate transformations and invariants within material media, from the above octonion composite physical quantities within material media.

\section{Galilean transformation}

Although the vector part of the octonion may vary, the scalar part of one octonion will remain unchanged still, in the rotational transformations of octonion coordinate systems. Making use of this property of octonion physical quantities, it is capable of inferring several classical invariants, conservation laws and continuity equations, in the electromagnetic and gravitational fields considering the contribution of material media.

\subsection{Law of mass conservation}

In the octonion space $\mathbb{O}$ , the octonion radius vector, $\mathbb{R} = i \textbf{i}_0 r_0 + \Sigma \textbf{i}_k r_k + k_{eg} ( i \textbf{I}_0 R_0 + \Sigma \textbf{I}_k R_k )$ , in the coordinate system $\alpha$ can be transformed into the octonion radius vector, $\mathbb{R}^\prime = i \textbf{i}_0^\prime r_0^\prime + \Sigma \textbf{i}_k^\prime r_k^\prime + k_{eg} ( i \textbf{I}_0^\prime R_0^\prime + \Sigma \textbf{I}_k^\prime R_k^\prime )$ , in the coordinate system $\beta$. In the rotational transformation of octonion coordinate system $\alpha$, the scalar part of one octonion keeps the same. Therefore,
\begin{eqnarray}
& r_0 = r_0^\prime  ~,
\end{eqnarray}
where $\textbf{i}_0^\prime = 1$.

Similarly, the octonion velocity $\mathbb{V}(v_j , V_j)$, linear momentum $\mathbb{P}^+(p_j^+ , P_j^+)$, angular momentum $\mathbb{L}^+(L_{1j}^+ , L_{2j}^+ , L_{1k}^{i+} , L_{2k}^{i+})$, torque $\mathbb{W}^+(W_{1j}^+ , W_{2j}^+ , W_{1j}^{i+} , W_{2j}^{i+})$ and force $\mathbb{N}^+(N_{1j}^+ , N_{2j}^+ , N_{1j}^{i+} , N_{2j}^{i+})$, in the coordinate system $\alpha$, can be transformed into the octonion velocity $\mathbb{V}^\prime(v_j^\prime , V_j^\prime)$, linear momentum $\mathbb{P}^{+\prime}(p_j^{+\prime} , P_j^{+\prime})$, angular momentum $\mathbb{L}^{+\prime}(L_{1j}^{+\prime} , L_{2j}^{+\prime} , L_{1k}^{i+\prime} , L_{2k}^{i+\prime})$, torque $\mathbb{W}^{+\prime}(W_{1j}^{+\prime} , W_{2j}^{+\prime} , W_{1j}^{i+\prime} , W_{2j}^{i+\prime})$ and force $\mathbb{N}^{+\prime}(N_{1j}^{+\prime} , N_{2j}^{+\prime} , N_{1j}^{i+\prime} , N_{2j}^{i+\prime})$ in the coordinate system $\beta$, respectively (see Ref.[28]). So there are,
\begin{eqnarray}
& v_0 = v_0^\prime  ~,
\\
& p_0^+ = p_0^{+\prime}  ~,
\\
& L_{10}^+ = L_{10}^{+\prime}  ~,
\\
& W_{10}^+ = W_{10}^{+\prime}  ~,
\\
& W_{10}^{i+} = W_{10}^{i+\prime}  ~,
\\
& N_{10}^+ = N_{10}^{+\prime}  ~,
\\
& N_{10}^{i+} = N_{10}^{i+\prime}  ~.
\end{eqnarray}

The combination of the above invariants can be chosen as the basic postulates of some types of coordinate transformations in the octonion space $\mathbb{O}$ , including the basic postulates, Eqs.(1) and (2), of Galilean transformation (Table 2). Apparently, various combinations of these invariants can be applied as the basic postulates for different coordinate transformations, picturing a few diverse overviews of the physical world. The invariants in the Table 2 can be established simultaneously.

The Table 2 consists of some classical conserved quantities. (a) Galilean transformation. From Eqs.(1) and (2), we can select not only the speed of light but also the scalar part of octonion radius vector to be conserved simultaneously. It implies that the time is conserved. And this is the basic postulate of familiar Galilean transformation. (b) Conserved mass. From Eqs.(2) and (3), one can choose the speed of light and the scalar part of octonion linear momentum both are conserved simultaneously. That is, the gravitational mass, $( p_0^+ / v_0 )$, is conserved. (c) Conserved energy. From Eqs.(2) and (6), it is capable of selecting the speed of light and energy both are conserved simultaneously. That is, the equivalent mass, $( W_{10}^{i+} / v_0^2 )$, is conserved. Further, it is able to achieve other types of conserved quantities.

The above conserved quantities are relatively simple so they will be applied comparatively often, in particular, the basic postulates of Galilean transformation, law of mass conservation, law of energy conservation, and fluid continuity equation (see Ref.[28]). The rest of the conserved quantities are relatively unfamiliar and they are comparatively less applied. Obviously these invariants have an important impact on the theoretical analysis. The different combinations of invariants can give their respective physical scenarios, enabling the theoretical description more colorful.

It means that some different physical quantities can be utilized to describe the physical phenomena from different perspectives. (a) According to Eqs.(1) and (2), it is able to choose the coordinate system, $S_r ( r_1 , r_2 , r_3 )$, to describe relevant physical phenomena. (b) According to Eqs.(2) and (3), one may select a coordinate system, $S_p ( p_1^+ , p_2^+ , p_3^+ )$, to explore the relevant physical phenomena in the momentum space. The utility of linear momenta replaces that of spatial coordinates, in the momentum space. (c) According to Eq.(2) and Eqs.(4) to (8), we can choose some different types of physical quantities to research the physical phenomena.

It is worth noting that, when the gravitational mass is conserved according to Eq.(3), the term relevant to the electric charge is one variable vector in the octonion space $\mathbb{O}$ . In other words, in case the law of mass conservation is effective, the electric charge is unable to be conserved. That is, the law of charge conservation is not tenable in the octonion space $\mathbb{O}$ .

But the law of charge conservation will be effective, in the transformed octonion space $\mathbb{O}_u$ relevant to the octonion space $\mathbb{O}$ .

\begin{table}[h]
\caption{Some conservation laws, continuity equations, and the basic postulate of Galilean transformation, in the octonion space $\mathbb{O}$ for the gravitational and electromagnetic fields. They can be derived from the invariants, when the speed of light is constant and the contribution of material media is considered.}
\begin{ruledtabular}
\begin{tabular}{lll}
speed of light~~~~~~              &   invariant                                                            &  conservation law                             \\
\hline
$v_0 = v_0^\prime$                &   $r_0 = r_0^\prime$                                                   &  Galilean transformation                      \\
                                  &   ~~~$t = t^\prime$                                                    &  ~~~conserved time                            \\
$v_0 = v_0^\prime$                &   $p_0^+ = p_0^{+\prime}$                                              &  ?                                            \\
                                  &   ~~~$m^+ = m^{+\prime}$                                               &  ~~~law of (gravitational) mass conservation  \\
$v_0 = v_0^\prime$                &   $L_{10}^+ = L_{10}^{+\prime}$                                        &  ?                                            \\
$v_0 = v_0^\prime$                &   $W_{10}^+ = W_{10}^{+\prime}$                                        &  ?                                            \\
$v_0 = v_0^\prime$                &   $W_{10}^{i+} = W_{10}^{i+\prime}$                                    &  law of energy conservation                   \\
                                  &   ~~~$W_{10}^{i+} / v_0^2 = W_{10}^{i+\prime} / v_0^{\prime2} $~~~~~~  &  ~~~law of (equivalent) mass conservation     \\
$v_0 = v_0^\prime$                &   $N_{10}^+ = N_{10}^{+\prime}$                                        &  (conserved) fluid continuity equation        \\
$v_0 = v_0^\prime$                &   $N_{10}^{i+} = N_{10}^{i+\prime}$                                    &  (conserved) torque continuity equation       \\
\end{tabular}
\end{ruledtabular}
\end{table}

\subsection{Law of charge conservation}

In the octonion space, if we multiply the basis vector, $i \textbf{I}_0$ , by the octonion radius vector, $\mathbb{R}$ , from the left, it is able to achieve one new octonion radius vector, $\mathbb{R}_u = i \textbf{I}_0 \circ \mathbb{R}$ . The latter can be considered as one octonion physical quantity in the transformed octonion space $\mathbb{O}_u$, that is, $\mathbb{R}_u = i \textbf{I}_0 \circ ( k_{eg} \mathbb{R}_e + \mathbb{R}_g )$. Apparently, the octonion radius vector $\mathbb{R}_u$ is independent of the octonion radius vector $\mathbb{R}$ , in particular the sequence of basis vectors or coordinate values.

In the transformed octonion space $\mathbb{O}_u$ , the octonion radius vector, $\mathbb{R}_u = k_{eg} ( R_0 + i \Sigma \textbf{i}_k R_k ) - ( \textbf{I}_0 r_0 + i \Sigma \textbf{I}_k r_k )$, in the coordinate system $\zeta$ can be transformed into the octonion radius vector, $\mathbb{R}_u^\prime = k_{eg} ( R_0^\prime + i \Sigma \textbf{i}_k^\prime R_k^\prime ) - ( \textbf{I}_0^\prime r_0^\prime + i \Sigma \textbf{I}_k^\prime r_k^\prime )$, in the coordinate system $\eta$. In the rotational transformation of octonion coordinate system $\zeta$, the scalar part of the octonion radius vector keeps the same. As a result,
\begin{eqnarray}
& R_0 = R_0^\prime  ~.
\end{eqnarray}

In a similar way, the octonion velocity $\mathbb{V}_u (v_j , V_j)$, linear momentum $\mathbb{P}^+_u (p_j^+ , P_j^+)$, angular momentum $\mathbb{L}^+_u (L_{1j}^+ , L_{2j}^+ , L_{1k}^{i+} , L_{2k}^{i+})$, torque $\mathbb{W}^+_u (W_{1j}^+ , W_{2j}^+ , W_{1j}^{i+} , W_{2j}^{i+})$ and force $\mathbb{N}^+_u (N_{1j}^+ , N_{2j}^+ , N_{1j}^{i+} , N_{2j}^{i+})$, in the coordinate system $\zeta$ , can be transformed into the octonion velocity $\mathbb{V}_u^\prime (v_j^\prime , V_j^\prime)$, linear momentum $\mathbb{P}_u^{+\prime} (p_j^{+\prime} , P_j^{+\prime})$, angular momentum $\mathbb{L}_u^{+\prime} (L_{1j}^{+\prime} , L_{2j}^{+\prime} , L_{1k}^{i+\prime} , L_{2k}^{i+\prime})$, torque $\mathbb{W}_u^{+\prime} (W_{1j}^{+\prime} , W_{2j}^{+\prime} , W_{1j}^{i+\prime} , W_{2j}^{i+\prime})$ and force $\mathbb{N}_u^{+\prime} (N_{1j}^{+\prime} , N_{2j}^{+\prime} , N_{1j}^{i+\prime} , N_{2j}^{i+\prime})$ in the coordinate system $\eta$, respectively (see Ref.[28]). So there are,
\begin{eqnarray}
& V_0 = V_0^\prime  ~,
\\
& P_0^+ = P_0^{+\prime}  ~,
\\
& L_{20}^+ = L_{20}^{+\prime}  ~,
\\
& W_{20}^+ = W_{20}^{+\prime}  ~,
\\
& W_{20}^{i+} = W_{20}^{i+\prime}  ~,
\\
& N_{20}^+ = N_{20}^{+\prime}  ~,
\\
& N_{20}^{i+} = N_{20}^{i+\prime}  ~.
\end{eqnarray}

The combination of these invariants can be chosen as the basic postulates of a few coordinate transformations in the transformed octonion space $\mathbb{O}_u$ . Obviously, various combinations of these invariants can be applied as the basic postulates for different types of coordinate transformations, describing several diverse overviews of the physical world (Table 3). The invariants in the Table 3 can be established simultaneously.

The Table 3 covers some conserved quantities. (a) Conserved electric charge. Choosing the two equations, Eqs.(10) and (11), shows that the electric charge, $( P_0^+ / V_0 )$, is conserved. (b) Conserved equivalent charge. The selection of two equations, Eqs.(10) and (14), states that the equivalent charge, $( W_{20}^{i+} / V_0^2 )$, is conserved. (c) Simultaneity. Selecting the conservation of second-speed of light, Eq.(10), means that the speed of light, $v_0$ , is not conserved, that is, Eq.(2) is unable to effective in the transformed octonion space $\mathbb{O}_u$ .

According to Eq.(10) and Eqs.(15) to (16), it is able to choose different types of coordinate systems to study the physical phenomena from multiple perspectives. These conserved quantities are relatively simple so they can be utilized comparatively often, in particular, the law of charge conservation and current continuity equation (see Ref.[28]). Although the rest of the conserved quantities are relatively strange, they also have an impact on the theoretical analysis.

It is rather remarkable that the transformed octonion space $\mathbb{O}_u$ is distinct to the octonion space $\mathbb{O}$ , so the invariants in the Table 3 cannot be established with those in the Table 2 simultaneously. In the transformed octonion space $\mathbb{O}_u$ , when the electric charge is conserved according to Eq.(11), the term relevant to the mass is one variable vector in the transformed octonion space $\mathbb{O}_u$ . In other words, in case the law of charge conservation is effective, the mass must not be conserved. That is, the law of mass conservation is not effective in the transformed octonion space $\mathbb{O}_u$ .

The norm of octonion radius vector is one scalar also, and remains unchanged in the rotational transformations of octonion coordinate systems. Consequently, the above research methods can be extended to the norm of octonion radius vector and so forth, exploring the basic postulate of Lorentz transformation, and continuity equations and others.

\begin{table}[h]
\caption{Some conservation laws, continuity equations, and the basic postulate of coordinate transformation, in the transformed octonion space $\mathbb{O}_u$ for the gravitational and electromagnetic fields. They can be derived from the invariants, when the second-speed of light is constant while the contribution of material media is considered.}
\begin{ruledtabular}
\begin{tabular}{lll}
second-speed of light~~~~~~           &   invariant                                                            &  conservation law                                \\
\hline
$V_0 = V_0^\prime$                    &   $R_0 = R_0^\prime$                                                   &  second-Galilean transformation                  \\
                                      &   ~~~$T = T^\prime$                                                    &  ~~~conserved second-time                        \\
$V_0 = V_0^\prime$                    &   $P_0^+ = P_0^{+\prime}$                                              &  ?                                               \\
                                      &   ~~~$q^+ = q^{+\prime}$                                               &  ~~~law of charge conservation                   \\
$V_0 = V_0^\prime$                    &   $L_{20}^+ = L_{20}^{+\prime}$                                        &  ?                                               \\
$V_0 = V_0^\prime$                    &   $W_{20}^+ = W_{20}^{+\prime}$                                        &  ?                                               \\
$V_0 = V_0^\prime$                    &   $W_{20}^{i+} = W_{20}^{i+\prime}$                                    &  law of second-energy conservation               \\
                                      &   ~~~$W_{20}^{i+} / V_0^2 = W_{20}^{i+\prime} / V_0^{\prime2} $~~~~~~  &  ~~~law of (equivalent) charge conservation      \\
$V_0 = V_0^\prime$                    &   $N_{20}^+ = N_{20}^{+\prime}$                                        &  (conserved) current continuity equation         \\
$V_0 = V_0^\prime$                    &   $N_{20}^{i+} = N_{20}^{i+\prime}$                                    &  (conserved) second-torque continuity equation   \\
\end{tabular}
\end{ruledtabular}
\end{table}

\section{Lorentz transformation}

The norm of an octonion physical quantity keeps the same, in the rotational transformations of octonion coordinate systems. Making use of this property of octonion physical quantities, it is capable of inferring some classical invariants, including the conservation laws and continuity equations relevant to the norm of octonion physical quantities, in the electromagnetic and gravitational fields considering the contribution of material media.

\subsection{Octonion space $\mathbb{O}$}

In the octonion space $\mathbb{O}$ , the octonion radius vector $\mathbb{R}$ in the coordinate system $\alpha$ can be transformed into the octonion radius vector $\mathbb{O}^\prime$ in the coordinate system $\beta$. In the rotational transformation of octonion coordinate system $\alpha$, the norm of octonion radius vector keeps the same. Therefore,
\begin{eqnarray}
& \mathbb{R} \circ \mathbb{R}^\ast = \mathbb{R}^\prime \circ \mathbb{R}^{\prime \ast} ~,
\end{eqnarray}
or
\begin{eqnarray}
& r_0^2 - \Sigma r_k^2 - k_{eg}^2 \Sigma R_j^2 = r_0^{\prime 2} - \Sigma r_k^{\prime 2} - k_{eg}^2 \Sigma R_j^{\prime 2} ~ .
\end{eqnarray}

Similarly, the octonion linear momentum $\mathbb{P}^+$ , angular momentum $\mathbb{L}^+$ , torque $\mathbb{W}^+$ and force $\mathbb{N}^+$ , in the coordinate system $\alpha$ , can be transformed into the octonion linear momentum $\mathbb{P}^{+\prime}$ , angular momentum $\mathbb{L}^{+\prime}$ , torque $\mathbb{W}^{+\prime}$ and force $\mathbb{N}^{+\prime}$ in the coordinate system $\beta$, respectively. In the rotational transformation of octonion coordinate system $\alpha$, each of these norms of octonion physical quantities keeps the same. So there are,
\begin{eqnarray}
&& p_0^{+ 2} - \Sigma p_k^{+ 2} - k_{eg}^2 \Sigma P_j^{+ 2} = p_0^{+ \prime 2} - \Sigma p_k^{+ \prime 2} - k_{eg}^2 \Sigma P_j^{+ \prime 2} ~,
\\
&& \mathbb{L}^+ \circ \mathbb{L}^{+ \ast} = \mathbb{L}^{+ \prime} \circ (\mathbb{L}^{+ \prime})^\ast ~,
\\
&& \mathbb{W}^+ \circ \mathbb{W}^{+ \ast} = \mathbb{W}^{+ \prime} \circ (\mathbb{W}^{+ \prime})^\ast ~,
\\
&& \mathbb{N}^+ \circ \mathbb{N}^{+ \ast} = \mathbb{N}^{+ \prime} \circ (\mathbb{N}^{+ \prime})^\ast ~.
\end{eqnarray}

The combination of the above invariants is capable of deducing the basic postulates of several coordinate transformations in the octonion space $\mathbb{O}$ , including the Lorentz transformation, Eqs.(17) and (2). Apparently, other combinations of the above invariants can be chosen as the basic postulates for different coordinate transformations, describing a few diverse overviews of the physical world (Table 4). The invariants in the Table 4 can be effective simultaneously.

The Table 4 includes a few classical conserved quantities. (a) Lorentz transformation. From Eqs.(17) and (2), one may select not only the speed of light but also the norm of octonion radius vector to be conserved simultaneously. It implies that the time is not conserved. And this is the basic postulate of familiar Lorentz transformation. (b) Non-conserved mass. From Eqs.(2) and (19), it is able to choose the speed of light and the norm of octonion linear momentum both are conserved simultaneously. That is, the gravitational mass, $( p_0^+ / v_0 )$, is not conserved. (c) Non-conserved energy. From Eqs.(2) and (21), it is capable of selecting the speed of light and the norm of octonion torque both are conserved simultaneously. That is, the energy term, $W_{10}^{i+}$ , is not conserved. Either the equivalent mass, $( W_{10}^{i+} / v_0^2 )$, is not conserved. (d) Non-conserved continuity equations. From Eqs.(2) and (22), we may choose not only the speed of light but also the norm of octonion force to be conserved simultaneously. It means that the continuity equations are not invariable nor conserved.

These conserved quantities are relatively simple and often used, in particular, the basic postulate of Lorentz transformation. The remaining conserved quantities are relatively unfamiliar, and they are comparatively rarely used. The combinations of remaining conserved quantities relevant to the norms have an important impact on the theoretical analysis also. Obviously, choosing the combination of different invariants as the basic postulate is able to achieve different physical scenarios.

\begin{table}[h]
\caption{Several conservation laws, continuity equations, and the basic postulate of Lorentz transformation are dependent on the norms of physical quantities, in the octonion space $\mathbb{O}$ for the gravitational and electromagnetic fields. They can be derived from the invariants relevant to the norms of physical quantities, when the speed of light is constant while the contribution of material media is considered.}
\begin{ruledtabular}
\begin{tabular}{lll}
speed of light~~~~~      & invariant relevant to the norm                                                                 &  conservation law                  \\
\hline
$v_0 = v_0^\prime$       & $r_0^2 - \Sigma r_k^2 - k_{eg}^2 \Sigma R_j^2 = r_0^{\prime 2} - \Sigma r_k^{\prime 2} - k_{eg}^2 \Sigma R_j^{\prime 2}$
                                                                                                                          & Lorentz transformation             \\
$v_0 = v_0^\prime$       & $p_0^{+ 2} - \Sigma p_k^{+ 2} - k_{eg}^2 \Sigma P_j^{+ 2} = p_0^{+ \prime 2} - \Sigma p_k^{+ \prime 2} - k_{eg}^2 \Sigma P_j^{+ \prime 2}$
                                                                                                                          & norm of octonion linear momentum   \\

                         & ~~~$(p_0^+ / v_0)^2 - \Sigma (p_k^+ / v_0)^2 - k_{eg}^2 \Sigma (P_j^+ / v_0)^2$   &  \\
                         & ~~~~~~$= (p_0^{+ \prime} / v_0^\prime)^2 - \Sigma (p_k^{+ \prime} / v_0^\prime)^2 - k_{eg}^2 \Sigma (P_j^{+ \prime} / v_0^\prime)^2$
                                                                                                                          & ~~~norm of mass                    \\
$v_0 = v_0^\prime$       & $\mathbb{L}^+ \circ \mathbb{L}^{+ \ast} = \mathbb{L}^{+ \prime} \circ (\mathbb{L}^{+ \prime})^\ast$
                                                                                                                          & norm of octonion angular momentum  \\
$v_0 = v_0^\prime$       & $\mathbb{W}^+ \circ \mathbb{W}^{+ \ast} = \mathbb{W}^{+ \prime} \circ (\mathbb{W}^{+ \prime})^\ast$
                                                                                                                          & norm of octonion torque            \\
                         & ~~~$(\mathbb{W}^+ / v_0^2) \circ (\mathbb{W}^+ / v_0^2)^\ast
                               = (\mathbb{W}^{+ \prime} / v_0^{\prime 2} ) \circ (\mathbb{W}^{+ \prime} / v_0^{\prime 2} )^\ast$~~~~
                                                                                                                          & ~~~norm of equivalent mass         \\
$v_0 = v_0^\prime$       & $\mathbb{N}^+ \circ \mathbb{N}^{+ \ast} = \mathbb{N}^{+ \prime} \circ (\mathbb{N}^{+ \prime})^\ast$
                                                                                                                          & norm of octonion force             \\
\end{tabular}
\end{ruledtabular}
\end{table}

\subsection{Octonion space $\mathbb{O}_u$}

In the transformed octonion space $\mathbb{O}_u$ , the octonion radius vector $\mathbb{R}_u$ in the coordinate system $\zeta$ can be transformed into the octonion radius vector $\mathbb{R}_u^\prime$ in the coordinate system $\eta$. In the rotational transformation of octonion coordinate system $\zeta$, the norm of octonion radius vector $\mathbb{R}_u$ keeps the same. It is easy to find that the norm of octonion radius vector $\mathbb{R}_u$ , in the transformed octonion space $\mathbb{O}_u$ , is identical to that of octonion radius vector $\mathbb{R}$, in the octonion space $\mathbb{O}$ .

Similarly, the norm of octonion linear momentum $\mathbb{P}_u^+$ , angular momentum $\mathbb{L}_u^+$ , torque $\mathbb{W}_u^+$ and force $\mathbb{N}_u^+$ , in the transformed octonion space $\mathbb{O}_u$, will be identical to that of octonion linear momentum $\mathbb{P}^+$ , angular momentum $\mathbb{L}^+$ , torque $\mathbb{W}^+$ and force $\mathbb{N}^+$ in the octonion space $\mathbb{O}$, respectively.

The combination of the above invariants can give some coordinate transformations, in the transformed octonion space $\mathbb{O}_u$ . Apparently, a few combinations of different invariants can deduce other types of coordinate transformations, depicting several different physical scenarios (Table 5). And that the invariants in the Table 5 can be effective simultaneously.

The Table 5 includes some conserved quantities. (a) Non-conserved electric charge. From Eqs.(10) and (19), it is able to choose the second-speed of light and the norm of octonion linear momentum both are conserved simultaneously. That is, the electric charge, $( P_0^+ / V_0 )$, is not conserved. (b) Non-conserved equivalent charge. From Eqs.(10) and (21), it is capable of selecting the second-speed of light and the norm of octonion torque both are conserved simultaneously. That is, the second-energy term, $W_{20}^{i+}$ , is not conserved. Either the equivalent charge, $( W_{20}^{i+} / V_0^2 )$, is not conserved. (c) Non-conserved continuity equations. From Eqs.(10) and (22), we may choose not only the second-speed of light but also the norm of octonion force to be conserved simultaneously. It means that the continuity equations are not invariable nor conserved. The different combinations of the remaining conserved quantities relevant to the norms have an important influence on the theoretical analysis.

The norm of octonion velocity is one scalar too, and keeps the same under the rotational transformations of octonion coordinate systems. So the above research methods will be extended to the norm of octonion velocity and others, researching the basic postulates of several coordinate transformations, and continuity equations and others.

\begin{table}[h]
\caption{Some conservation laws, continuity equations, and the basic postulates of coordinate transformations are dependent on the norms of physical quantities, in the transformed octonion space $\mathbb{O}_u$ for the gravitational and electromagnetic fields. They can be derived from the invariants relevant to the norms of physical quantities, when the second-speed of light is constant while the contribution of material media is considered.}
\begin{ruledtabular}
\begin{tabular}{lll}
second-speed of light~~~ & invariant relevant to the norm                                                                 &  conservation law                  \\
\hline
$V_0 = V_0^\prime$       & $r_0^2 - \Sigma r_k^2 - k_{eg}^2 \Sigma R_j^2 = r_0^{\prime 2} - \Sigma r_k^{\prime 2} - k_{eg}^2 \Sigma R_j^{\prime 2}$
                                                                                                                          & second-Lorentz transformation      \\
$V_0 = V_0^\prime$       & $p_0^{+ 2} - \Sigma p_k^{+ 2} - k_{eg}^2 \Sigma P_j^{+ 2} = p_0^{+ \prime 2} - \Sigma p_k^{+ \prime 2} - k_{eg}^2 \Sigma P_j^{+ \prime 2}$
                                                                                                                          & norm of octonion linear momentum   \\
                         & ~~~$(p_0^+ / V_0)^2 - \Sigma (p_k^+ / V_0)^2 - k_{eg}^2 \Sigma (P_j^+ / V_0)^2$                &                                    \\
                         & ~~~~~~$= (p_0^{+ \prime} / V_0^\prime)^2 - \Sigma (p_k^{+ \prime} / V_0^\prime)^2 - k_{eg}^2 \Sigma (P_j^{+ \prime} / V_0^\prime)^2$
                                                                                                                          & ~~~norm of electric charge         \\
$V_0 = V_0^\prime$       & $\mathbb{L}^+ \circ \mathbb{L}^{+ \ast} = \mathbb{L}^{+ \prime} \circ (\mathbb{L}^{+ \prime})^\ast$
                                                                                                                          & norm of octonion angular momentum  \\
$V_0 = V_0^\prime$       & $\mathbb{W}^+ \circ \mathbb{W}^{+ \ast} = \mathbb{W}^{+ \prime} \circ (\mathbb{W}^{+ \prime})^\ast$
                                                                                                                          & norm of octonion torque            \\
                         & ~~~$(\mathbb{W}^+ / V_0^2) \circ (\mathbb{W}^+ / V_0^2)^\ast
                               = (\mathbb{W}^{+ \prime} / V_0^{\prime 2} ) \circ (\mathbb{W}^{+ \prime} / V_0^{\prime 2} )^\ast$
                                                                                                                          & ~~~norm of equivalent charge       \\
$V_0 = V_0^\prime$       & $\mathbb{N}^+ \circ \mathbb{N}^{+ \ast} = \mathbb{N}^{+ \prime} \circ (\mathbb{N}^{+ \prime})^\ast$
                                                                                                                          & norm of octonion force             \\
\end{tabular}
\end{ruledtabular}
\end{table}

\section{Norm of octonion velocity}

The norm of octonion velocity remains unchanged, in the rotational transformations of octonion coordinate systems. According to the octonion property, it is able to achieve a few invariants relevant to the norm of octonion velocity, including the conservation laws and continuity equations, in the electromagnetic and gravitational fields considering the contribution of material media.

In the octonion space $\mathbb{O}$ , the octonion velocity $\mathbb{V}$ in the coordinate system $\alpha$ can be transformed into the octonion velocity $\mathbb{V}^\prime$ in the coordinate system $\beta$. In the rotational transformation of octonion coordinate system $\alpha$, the norm of octonion velocity keeps the same. Therefore,
\begin{eqnarray}
& \mathbb{V} \circ \mathbb{V}^\ast = \mathbb{V}^\prime \circ \mathbb{V}^{\prime \ast} ~,
\end{eqnarray}
or
\begin{eqnarray}
& v_0^2 - \Sigma v_k^2 - k_{eg}^2 \Sigma V_j^2 = v_0^{\prime 2} - \Sigma v_k^{\prime 2} - k_{eg}^2 \Sigma V_j^{\prime 2} ~ .
\end{eqnarray}

Similarly, each of the norms of octonion radius vector $\mathbb{R}$ , linear momentum $\mathbb{P}^+$ , angular momentum $\mathbb{L}^+$ , torque $\mathbb{W}^+$ and force $\mathbb{N}^+$ keeps the same, in the rotational transformation of octonion coordinate system (see Tables 4 and 5). The selection of two invariants, Eqs.(24) and (19), will deduce the transformation of the norm of octonion linear momentum.

In the octonion space $\mathbb{O}$ , Eq.(24) can be simplified into,
\begin{eqnarray}
v_{0(2)} = v_{0(2)}^\prime  ~,
\end{eqnarray}
where $v_{0(2)}^2 =  v_0^2 - \Sigma v_k^2 - k_{eg}^2 \Sigma V_j^2 $ , $v_{0(2)}^{\prime 2} = v_0^{\prime 2} - \Sigma v_k^{\prime 2} - k_{eg}^2 \Sigma V_j^{\prime 2}$ .

The above means that we may select the speed of light, $v_{0(2)}$ , to be invariable, in the octonion space $\mathbb{O}$ . From two equations, Eqs.(19) and (25) , there is the transformation of norm of mass,
\begin{eqnarray}
&& (p_0^+ / v_{0(2)})^2 - \Sigma (p_k^+ / v_{0(2)})^2 - k_{eg}^2 \Sigma (P_j^+ / v_{0(2)})^2
\nonumber
\\
= && (p_0^{+ \prime} / v_{0(2)}^\prime)^2 - \Sigma (p_k^{+ \prime} / v_{0(2)}^\prime)^2 - k_{eg}^2 \Sigma (P_j^{+ \prime} / v_{0(2)}^\prime)^2  ~,
\end{eqnarray}

In the transformed octonion space $\mathbb{O}_u$ , Eq.(24) is also reduced into,
\begin{eqnarray}
V_{0(2)} = V_{0(2)}^\prime  ~,
\end{eqnarray}
where $- k_{eg}^2 V_{0(2)}^2 =  v_0^2 - \Sigma v_k^2 - k_{eg}^2 \Sigma V_j^2 $ , $- k_{eg}^2 V_{0(2)}^{\prime 2} = v_0^{\prime 2} - \Sigma v_k^{\prime 2} - k_{eg}^2 \Sigma V_j^{\prime 2}$ .

The above implies that it is able to choose the second-speed of light, $V_{0(2)}$ , to be invariable, in the transformed octonion space $\mathbb{O}_u$ . From two equations, Eqs.(19) and (27) , there is the transformation of norm of electric charge,
\begin{eqnarray}
&& (p_0^+ / V_{0(2)})^2 - \Sigma (p_k^+ / V_{0(2)})^2 - k_{eg}^2 \Sigma (P_j^+ / V_{0(2)})^2
\nonumber
\\
= && (p_0^{+ \prime} / V_{0(2)}^\prime)^2 - \Sigma (p_k^{+ \prime} / V_{0(2)}^\prime)^2 - k_{eg}^2 \Sigma (P_j^{+ \prime} / V_{0(2)}^\prime)^2  ~,
\end{eqnarray}

In terms of the norm of octonion torque, it is able to achieve a few similar inferences (Table 6). The selection of two equations, Eqs.(25) and (21), will deduce the transformation of the norm of equivalent mass, when the speed of light, $v_{0(2)}$ , is invariable in the octonion space $\mathbb{O}$ . Next, choosing the two equations, Eqs.(27) and (21), will infer the transformation of the norm of equivalent charge, when the second-speed of light, $V_{0(2)}$ , is invariable in the transformed octonion space $\mathbb{O}_u$ . It is a remarkable fact that some invariants will become the variable quantities in the Tables 2 and 3, after Eq.(24) replaces Eq.(2) or Eq.(10). In other words, these physical quantities are no longer conserved quantities. From the perspective of octonion spaces, each conserved quantity has several preconditions, in particular, the time and speed of light that are often used.

Consequently, the ratio of charge to mass must be variable, that is, it is not an invariant \cite{mpyle}. Moreover, the reason for the variation of the charge-to-mass ratio is quite complicated. (a) According to the invariants in the Tables 2 and 3, it is able to deduce that the ratio of charge to mass is not an invariant. (b) From the invariants in the Tables 4 and 5, it is capable of inferring that the ratio of charge to mass is not an invariant. (c) One can derive that the ratio of charge to mass is not an invariant, from the invariants in the Table 6. (d) A part of the octonion field potential will also exert an impact on the speed of light, $v_0$ , and second-speed of light, $V_0$ , resulting in the variation of the ratio of charge to mass.

The above research shows that, a few combinations of invariants can be selected as the basic postulates of some coordinate transformations, from different focuses of considerations. They can propose several different theoretical explanations of physics, describing various types of physical phenomena, in particular the varying speed of light and variable optical index and others within the optical materials.

\begin{table}[h]
\caption{Some conservation laws, continuity equations, and the basic postulates of coordinate transformations are dependent on the norms of physical quantities, in the octonion space for the gravitational and electromagnetic fields. They can be derived from the invariants relevant to the norms of physical quantities, when the speed of light is not constant while the contribution of material media is considered.}
\begin{ruledtabular}
\begin{tabular}{lll}
first invariant          & second invariant                                                                               &  conservation law                      \\
\hline
$\mathbb{V} \circ \mathbb{V}^\ast = \mathbb{V}^\prime \circ \mathbb{V}^{\prime \ast}$~~
                         & $r_0^2 - \Sigma r_k^2 - k_{eg}^2 \Sigma R_j^2 = r_0^{\prime 2} - \Sigma r_k^{\prime 2} - k_{eg}^2 \Sigma R_j^{\prime 2}$
                                                                                                                          & norm of octonion radius vector        \\
$\mathbb{V} \circ \mathbb{V}^\ast = \mathbb{V}^\prime \circ \mathbb{V}^{\prime \ast}$
                         & $p_0^{+ 2} - \Sigma p_k^{+ 2} - k_{eg}^2 \Sigma P_j^{+ 2} = p_0^{+ \prime 2} - \Sigma p_k^{+ \prime 2} - k_{eg}^2 \Sigma P_j^{+ \prime 2}$
                                                                                                                          & norm of octonion linear momentum      \\
~~~$v_{0(2)} = v_{0(2)}^\prime$
                         & ~~~$( p_0^+ / v_{0(2)} )^2 - \Sigma (p_k^+ / v_{0(2)} )^2 - k_{eg}^2 \Sigma (P_j^+ / v_{0(2)} )^2$   &                                 \\
                         & ~~~~~~$= (p_0^{+ \prime} / v_{0(2)}^\prime )^2 - \Sigma (p_k^{+ \prime} / v_{0(2)}^\prime )^2 - k_{eg}^2 \Sigma (P_j^{+ \prime} / v_{0(2)}^\prime )^2$
                                                                                                                          & ~~~norm of mass                        \\
~~~$V_{0(2)} = V_{0(2)}^\prime$
                         & ~~~$( p_0^+ / V_{0(2)} )^2 - \Sigma (p_k^+ / V_{0(2)} )^2 - k_{eg}^2 \Sigma (P_j^+ / V_{0(2)} )^2$   &                                  \\
                         & ~~~~~~$= (p_0^{+ \prime} / V_{0(2)}^\prime )^2 - \Sigma (p_k^{+ \prime} / V_{0(2)}^\prime )^2 - k_{eg}^2 \Sigma (P_j^{+ \prime} / V_{0(2)}^\prime )^2$
                                                                                                                          & ~~~norm of charge                      \\
$\mathbb{V} \circ \mathbb{V}^\ast = \mathbb{V}^\prime \circ \mathbb{V}^{\prime \ast}$
                         & $\mathbb{L}^+ \circ \mathbb{L}^{+ \ast} = \mathbb{L}^{+ \prime} \circ (\mathbb{L}^{+ \prime})^\ast$
                                                                                                                          & norm of octonion angular momentum     \\
$\mathbb{V} \circ \mathbb{V}^\ast = \mathbb{V}^\prime \circ \mathbb{V}^{\prime \ast}$
                         & $\mathbb{W}^+ \circ \mathbb{W}^{+ \ast} = \mathbb{W}^{+ \prime} \circ (\mathbb{W}^{+ \prime})^\ast$
                                                                                                                          & norm of octonion torque                \\
~~~$v_{0(2)} = v_{0(2)}^\prime$
                         & ~~~$(\mathbb{W}^+ / v_{0(2)}^2) \circ (\mathbb{W}^+ / v_{0(2)}^2)^\ast$                        &                                        \\
                         & ~~~~~~$= (\mathbb{W}^{+ \prime} / v_{0(2)}^{\prime 2} ) \circ (\mathbb{W}^{+ \prime} / v_{0(2)}^{\prime 2} )^\ast$
                                                                                                                          & ~~~norm of equivalent mass             \\
~~~$V_{0(2)} = V_{0(2)}^\prime$
                        & ~~~$(\mathbb{W}^+ / V_{0(2)}^2) \circ (\mathbb{W}^+ / V_{0(2)}^2)^\ast$                         &                                        \\
                        & ~~~~~~$= (\mathbb{W}^{+ \prime} / V_{0(2)}^{\prime 2} ) \circ (\mathbb{W}^{+ \prime} / V_{0(2)}^{\prime 2} )^\ast$
                                                                                                                          & ~~~norm of equivalent charge           \\
$\mathbb{V} \circ \mathbb{V}^\ast = \mathbb{V}^\prime \circ \mathbb{V}^{\prime \ast}$
                         & $\mathbb{N}^+ \circ \mathbb{N}^{+ \ast} = \mathbb{N}^{+ \prime} \circ (\mathbb{N}^{+ \prime})^\ast$
                                                                                                                          & norm of octonion force                 \\
\end{tabular}
\end{ruledtabular}
\end{table}

\section{Discussions}

The scalar part of an octonion will keep the same, in case the octonion coordinate system transforms rotationally. Making use of this octonion property, it is able to deduce some invariants of the Tables 2 to 6 in the octonion space, including the basic postulates of Galilean transformation, Lorentz transformation, and transformation of the norm of octonion velocity and so forth.

(1) Galilean transformation. The basic postulate of Galilean transformation and other invariants, in the Tables 2 and 3, play an important role in some cases. Meanwhile the contributions of invariants are negligible in the Tables 4 to 6. This is one of viewpoints about the coordinate transformations in the physics before the 20th century. It properly describes most of the physical phenomena encountered in the classical physics, in particular the cases of low-speed movement or low field strength.

(2) Lorentz transformation. The basic postulate of Lorentz transformation and other invariants, in the Tables 4 and 5, play a major role in certain circumstances. And the contributions of invariants can be neglected in the Tables 2, 3 and 6. This is one type of viewpoint about the relativity in the 20th century. It is capable of describing some special physical phenomena, in particular the high-speed motions, with the speed of light being a constant.

(3) Transformation of the norm. The basic postulates of the transformations, relevant to the norm of octonion velocity, and other invariants, in the Table 6, act as an important part in some situations. Meanwhile, it allows us to neglect the contributions of invariants in the Tables 2 to 5. This is one point of view in the paper. It is capable of depicting a few special physical phenomena, including some situations of varying speed of light and variable refractive indices.

In the octonion spaces, there are several preconditions for each conserved quantity in the field theories, in particular the time and the speed of light that are often used, from the perspective of octonion spaces. As a result, the physical scenarios you see are different from each other, from the perspective of different combinations of invariants. When some multiple combinations of invariants play an important role, they are able to explore a variety of physical phenomena. If each of these multiple combinations of invariants has a certain contribution, we can achieve the superposition of multiple physical scenarios. Consequently the physical phenomena that field theory can describe are much more colorful than ever before (Table 7).

Compared with the Tables 4 and 5, the Tables 2 and 3 are seized of more equations, enabling them to describe more detailed physical properties of physical phenomena. Consequently, these equations in the Tables 2 and 3 are easier to be widely applied as a theoretical description. On the other hand, compared with the Tables 4 and 5, the Table 6 is possessed of fewer equations, exploring several coarser physical properties of physical phenomena. Therefore, these equations in the Table 6 are utilized less as a theoretical statement. By comparing the three types of coordinate transformations, it can be found that the Galilean transformation may depict one larger number of invariant details. The Lorentz transformation will picture one small number of invariant details. The comparatively complicated coordinate transformations, relevant to the norm of octonion velocity, can only describe one smallest number of invariant details. Apparently, in the theoretical interpretation of physical phenomena, the equations in the Tables 2 and 3 must be preferred for the theoretical analysis of physical phenomena, in order to obtain as many details of theoretical narration as possible. This is an important reason why the equations in the Tables 2 and 3 can be widely used. Thereafter, the Tables 4 to 6 serve as the supplementary means for theoretical descriptions.

In the actual theoretical description of a large number of physical phenomena that occur simultaneously, the theoretical description will be based on the physical description of invariants and Galilean transformation in the Tables 2 and 3. For some cases, it is necessary to superimpose the physical description of invariants and Lorentz transformation in the Tables 4 and 5, affording an appropriate theoretical description of some simultaneous physical phenomena. And even the physical description of invariants and transformation of norms in the Table 6 should be superimposed, for a fewer cases. It enables the physical descriptions to describe more accurately a large number of simultaneous physical phenomena.

In other words, most physical events involve the Galilean transformation and rest invariants in the Tables 2 and 3, for the physical events that have been involved so far. Few physical events involve the Lorentz transformation and other invariants in the Tables 4 to 5. And that fewer physical events involve the transformation of norm of octonion velocity and other invariants in the Table 6.

In terms of the energy conservation, in case the Tables 2 and 3 can be considered as the special case of the Tables 4 and 5, respectively, the scope of application of the norm of octonion torque, Eq.(21), in the Table 4 will be much larger than that of the law of energy conservation, Eq.(5), in the Table 2.

Through comparison and analysis, it can be found that the theoretical explanation of some physical phenomena, relevant to the varying speed of light, is independent from the that in the octonion composite spaces \cite{weng3}. In the octonion composite spaces, the octonion composite radius vector is, $\mathbb{R}^+ = \mathbb{R} + k_{rx} \mathbb{X}$ . And the quaternion composite operator is, $\lozenge^+ = i \textbf{i}_0 \partial / \partial r_0^+ + \Sigma \textbf{i}_k \partial / \partial r_k^+$, and $r_j^+ = r_j + k_{rx} x_j$ .

As mentioned above, the invariants can be divided into two different groups, in the gravitational and electromagnetic fields. The first group of invariants is in the octonion space $\mathbb{O}$ , including the Tables 2 and 4. So these invariants can merely hold in the octonion space $\mathbb{O}$ . The second group of invariants is in the transformed octonion space $\mathbb{O}_u$ , including the Tables 3 and 5. As a result, these invariants can only be established in the transformed octonion space $\mathbb{O}_u$ . Apparently, the first group of invariants, in the octonion space $\mathbb{O}$ , cannot be established with the second group of invariants, in the transformed octonion space $\mathbb{O}_u$ , simultaneously.

\begin{table}[h]
\caption{In the octonion spaces for the electromagnetic and gravitational fields, the comparison of physical characteristics among several main coordinate transformations, including the basic postulates and mathematical difficulty as well as the number of equations.}
\begin{ruledtabular}
\begin{tabular}{lllll}
coordinate transformation       &  basic postulate         &  mathematical            &  number of                   &  octonion                           \\
                                &                          &  ~~~~~difficulty         &  ~~equations                 &  ~~~space                           \\
\hline
Galilean transformation         &  Eqs.(1) and (2)         &  simple                  &  many                        &  $\mathbb{O}$                       \\
second-Galilean transformation  &  Eqs.(9) and (10)	       &  simple                  &  many                        &  $\mathbb{O}_u$                     \\
Lorentz transformation          &  Eqs.(2) and (18)        &  complicated             &  few                         &  $\mathbb{O}$                       \\
second-Lorentz transformation   &  Eqs.(10) and (18)       &  complicated             &  few                         &  $\mathbb{O}_u$                     \\
transformation of norm          &  Eqs.(18) and            &  more                    &                              &                                     \\
~~~~~~~~~of octonion velocity   &   ~~~(25)/(27)           &  ~~complicated           &  lesser                      &  $\mathbb{O}$ or $\mathbb{O}_u$     \\
\end{tabular}
\end{ruledtabular}
\end{table}

\section{Conclusions}

J. C. Maxwell first utilized the algebra of quaternions to describe the physical properties of electromagnetic fields. This inspired the subsequent scholars to apply the quaternions to study the physical properties of gravitational fields. Nowadays, the scholars use the algebra of octonions to explore the physical properties of electromagnetic and gravitational fields simultaneously, including the octonion radius vector, velocity, field strength, field source, angular momentum, torque and force and others.

The scalar part of an octonion will remain unchanged, when the octonion coordinate system transforms rotationally. By means of this octonion property, it is able to deduce a few invariants, including the conservation laws, continuity equations, and the basic postulates of some coordinate transformations. The scalar parts of the octonion radius vector and velocity both remain unchanged, in the octonion coordinate system of rotational transformation. In the octonion space $\mathbb{O}$ , the two invariants can be selected as the basic postulates of Galilean transformation. In the octonion space, the Galilean transformation believes that both of the time and mass are invariants. Similarly, each of the scalar parts of octonion linear momentum and torque and so forth is an invariant also. The law of mass conservation and fluid continuity equation and others can be derived from these invariants and unvarying speed of light. In the transformed octonion space $\mathbb{O}_u$ , it is capable of inferring the law of charge conservation and current continuity equation and so forth. In the octonion space, the second-Galilean transformation deems that the electric charge is an invariant. Nevertheless, the transformed octonion space $\mathbb{O}_u$ is independent of the octonion space $\mathbb{O}$ . Consequently, the law of charge conservation and law of mass conservation cannot be established simultaneously. The current continuity equation and fluid continuity equation are unable to be effective simultaneously.

The norm of an octonion is the scalar also. The norm of the octonion will remain unchanged, when the octonion coordinate system transforms rotationally. Making use of this octonion property, it is capable of inferring several invariants relevant to the norms, including the conservation laws and the basic postulates of some coordinate transformations. In terms of the octonion radius vector, its norm keeps the same, in the octonion coordinate system of rotational transformation. In the octonion space $\mathbb{O}$ , this invariable norm and unvarying speed of light can be chosen as the basic postulates of Lorentz transformation. In the octonion space, the Lorentz transformation reckons that neither the time nor the mass is an invariant. Similarly, each of the norms of octonion linear momentum and torque and others is not an invariant either. The law of conservation relevant to the norm of mass and others can be derived from these invariants and unvarying speed of light. In the transformed octonion space $\mathbb{O}_u$ , it is able to conclude the law of conservation relevant to the norm of electric charge and so forth. In the octonion space, the second-Lorentz transformation considers that the electric charge is not an invariant. However, the law of conservation relevant to the norm of charge is unable to be established with the law of conservation relevant to the norm of mass simultaneously, because the transformed octonion space $\mathbb{O}_u$ is independent of the octonion space $\mathbb{O}$ .

The norm of octonion velocity is also a scalar and remains unchanged, when the octonion coordinate system transforms rotationally. From this octonion property, it is capable of deriving a few more complicated invariants related with the norms, including the conservation laws and the basic postulates of some coordinate transformations. In the octonion coordinate system of rotational transformation, the norms of the octonion radius vector and velocity both remain unchanged. If these two invariants are selected as the basic postulates, we may get a more complicated coordinate transformation related to the norm of octonion velocity. In the coordinate transformation, the time, mass, and electric charge and so forth are not all invariable.

In the octonion space $\mathbb{O}$ , compared with the number of equations related to the Lorentz transformation in the Tables 4 and 5, the Galilean transformation in the Tables 2 and 3 is relevant to more equations. The latter in the Tables 2 and 3 are provided with lower mathematical difficulty, describing one larger number of invariant details. Conversely, for the complicated coordinate transformation related to the norm of octonion velocity in the Table 6, the number of related equations is less than that of the Lorentz transformation in the Tables 4 and 5. The equations in the Table 6 take on higher mathematical difficulty, and they can merely investigate the invariant details more roughly. In other words, the equations relevant to the Galilean transformation are able to describe the maximum number of invariant details. The relevant case of Lorentz transformation involves the second largest number of invariant details. The equations of complicated coordinate transformation, relevant to the norm of octonion velocity, can merely explore the minimum number of invariant details. Further, there is also a similar situation in the transformed octonion space $\mathbb{O}_u$ as in the octonion space $\mathbb{O}$ . Consequently we must first choose the equations related to the Galilean transformation, in the Tables 2 and 3, to explain physical phenomena in the vast majority of cases.

It is noteworthy that this paper discusses only a few simple cases of invariants and physical properties of rotational transformations of coordinate systems in the octonion spaces, including the basic postulates of the Galilean transformation and Lorentz transformation. But it has clearly pointed out that, the basic postulates of Galilean transformation and Lorentz transformation both are invariants of the rotational transformation of the octonion coordinate systems. Choosing different combinations of invariants will infer the basic postulates of other types of coordinate transformations. Therefore, it can deduce the basic postulates of more complicated coordinate transformations, in particular several predictions related with the mass and electric charge. In the future studies, the paper plans to explore some more complicated coordinate transformations, verifying the relevant theoretical predictions in experiments.

\begin{acknowledgements}
The author thanks Mr. Heng-Lin Wang, one of their former colleagues, for the interest in this research and financial support. And the author is indebted to the anonymous referees for their valuable comments on the previous manuscripts. This project was supported partially by the National Natural Science Foundation of China under grant number 60677039.
\end{acknowledgements}

\section*{Conflict of interest}
The author has no conflicts to disclose.

\section*{Data Availability}
The data that support the finding of this study are available within this article.

{}

\end{document}